\documentclass[12pt,preprint]{aastex}

\def\lea{\mathrel{<\kern-1.0em\lower0.9ex\hbox{$\sim$}}}
\def\gea{\mathrel{>\kern-1.0em\lower0.9ex\hbox{$\sim$}}}

\slugcomment{}

\shorttitle{Deep M67 NIR CMD}
\shortauthors{Sarajedini et al.}

\begin{document}


\title{Deep 2MASS  Photometry of M67 and Calibration \\
of the Main Sequence J--K$_S$ Color Difference as an Age Indicator }

\author{Ata Sarajedini}
\affil{Department of Astronomy, University of Florida, Gainesville, FL 32611, USA}
\email{ata@astro.ufl.edu}

\author{Aaron Dotter}
\affil{Department of Physics and Astronomy, University of Victoria, Victoria, BC, Canada}
\email{dotter@uvic.ca}

\author{Allison Kirkpatrick}
\affil{Department of Astronomy, University of Florida, Gainesville, FL 32611, USA}
\email{allison@astro.ufl.edu}



\begin{abstract}
We present an analysis of Two Micron All Sky Survey (2MASS) calibration 
photometry of the old open cluster M67 (NGC 2682). The proper 
motion-cleaned color-magnitude
diagram (CMD) resulting from these data extends $\sim$3 magnitudes deeper
than one based on data from the point source catalog. The
CMD extends from above the helium-burning red clump to a faint limit
that is more than 7 magnitudes below the main sequence turnoff in the
K$_S$ band. After adopting a reddening of E(B--V) = 0.041 $\pm$ 0.004 
and a metal abundance of [Fe/H] = --0.009 $\pm$ 0.009 based on
a survey of published values, we fit the unevolved main sequence of 
M67 to field main sequence stars with 2MASS photometry and {\it Hipparcos}
parallaxes. This analysis yields distance moduli of
(m--M)$_{K_S}$ = 9.72 $\pm$ 0.05 and  
(m--M)$_0$ = 9.70 $\pm$ 0.05, which are consistent with published
values. We compare the theoretical isochrones 
of Girardi et al. and Dotter et al. to
the CMD of M67 and comment on the relative merits of each set of models. 
These comparisons suggest an age between 3.5 and 4.0 Gyr for M67. 
The depth of the M67 data make them ideal for the calibration
of a new age indicator that has recently been devised by Calamida et al.- 
the difference in
(J--K$_S$) color between the main sequence turnoff (TO) and the point on the 
lower main sequence where it turns down (TD) and becomes nearly vertical
[$\Delta$(J--K$_S$)$_{TD}^{TO}$]. Coupled with deep 2MASS photometry for three 
other open clusters, NGC 2516, M44, and NGC 6791, we calibrate $\Delta$(J--K$_S$) 
in terms of age and find 
$\Delta$(J--K$_S$)$_{TD}^{TO}$ = (3.017 $\pm$ 0.347) -- (0.259 $\pm$ 0.037)*Log Age (yrs).
\end{abstract}




\section{Introduction}

Infrared photometry of open clusters holds great
promise for many reasons. Most of the effort thus far in this area 
has been geared 
toward the study of embedded young star clusters in which the primary 
goal has been to investigate the star (and planet) forming properties of these 
systems. The infrared is a natural choice in this regard because of 
its ability to minimize dust obscuration ($A_{K} \sim 0.1 A_{V}$), 
which is often a problem in these objects as they are still 
enshrouded in their molecular clouds and preferentially located in the disk of 
the Milky Way.

An often overlooked use of infrared photometry is in the study of 
`normal' open clusters -- that is to say -- clusters that do not 
necessarily suffer from extreme levels of foreground extinction.
Detailed studies of these clusters in the near-IR are important for a number of
reasons. First, they can be used to calibrate the variation of features in
the color-magnitude diagram (CMD) with cluster properties such as
age, metallicity, and distance. Such a calibration provides a means
to determine these properties in clusters of interest using 
proxies in the CMD. Second, near-IR photometry
of low-reddening open clusters is important in order to test the 
color calibrations of theoretical isochrones. In cases where the
models perform poorly, these data provide an empirical set of
ridge lines that can be used to interpret the observations and 
indicate particular areas where the stellar models require improvement.
These tests are especially important when the properties of
stellar populations in distant galaxies are being investigated by
comparing population synthesis models to integrated photometry and 
spectroscopy. 

With regard to providing high quality, well-calibrated photometry
for low-reddening open clusters, the Two Micron All Sky Survey 
(2MASS, Skrutskie et al. 2006), is a significant resource. 
Because of its all-sky
coverage and moderate depth, the 2MASS Point Source Catalog
(PSC)
contains photometry for countless `normal' open cluster in the
JHK$_S$ passbands. Many studies have taken advantage of the
PSC in order to probe a variety of 
scientific questions (e.g. Grocholski \& Sarajedini 2002;
An et al. 2007, and references therein). A 2MASS resource that is less
frequently utilized is the large collection of pointed calibration
data taken to standardize the all-sky photometry. These are 
described in the work by Nikolaev et al. (2000) and consist of 
35 fields distributed in the northern and southern hemispheres
that are observed hundreds of times. One of these fields 
(No. 90067) contains the well-known, solar age and metallicity 
open cluster M67 (NGC 2682), which is the subject of the present
work. 

The paper is organized as follows. In the next section, we describe
the 2MASS calibration data that we plan to analyze. The
morphological features of the CMD are described in Sec. 3, while
the next section is devoted to a description of the main sequence
fitting method we employ to derive the distance to M67. Section 5
presents a comparison of the 2MASS near-IR CMD of M67 to two sets
of theoretical isochrones that are available in the 2MASS filter
set. We combine the deep M67 data with similar photometry for three 
other open clusters in Sec. 6 and investigate the validity of a new a
ge indicator recently unveiled in the literature. Finally, our
conclusions are summarized in Sec. 7.

\section{Observational Data}

The observations of M67 analyzed herein were taken as part of
the 2MASS calibration process. They
were obtained from the `Combined 2MASS
Calibration Scan' source list available from the 2MASS web 
site.~\footnote{http://www.ipac.caltech.edu/2mass/releases/allsky/doc/seca7\_4.html}
The 90067 calibration tile, which includes M67, contains 3771 point sources.
Of these, 3413 are flagged as genuine, well-measured stars (i.e. not artifacts
and not affected by artifacts); these are measured from averages of the best
images available, yielding a photometric limit that is 3 to 4 magnitudes deeper 
than the 2MASS point source catalog itself. 

\section{Color Magnitude Diagram}

The middle panel of Fig. 1 shows the color-magnitude diagram (CMD) 
of M67 from the deep 2MASS calibration observations. For 
comparison, the left panel displays the CMD from the 2MASS PSC 
over the same area. It is evident
from this comparison that the calibration data are not only deeper 
but also exhibit better-defined cluster principal sequences. In particular, 
below K$_S$$\sim$15, we see the main sequence (MS) becoming 
essentially vertical at $\sim$0.6M$_{\odot}$ where the J--K$_S$ color 
starts to become insensitive to effective temperature. Though it 
presumably exists in all star clusters, this downturn in the MS has 
only been observed in a few systems because it only appears 
in cool, dense MS stars whose absolute magnitudes are below the level
reached by most near-IR photometric surveys for objects with 
distance moduli of $\sim$10 or greater. The right panel of Fig. 1
shows the result of combining the 2MASS calibration photometry
with the proper motion data of Yadav et al. (2008) and including only
stars with a probability of membership greater than 20\%. This cleans
up the M67 CMD considerably clearly defining its principal sequences.
For the remainder of the paper, we will use these proper motion selected
data for our analysis. 

The downturn in the MS is caused by collisionally induced absorption (CIA)
of H$_2$ molecules that begin to appear in the atmospheres of the high gravity 
main sequence stars at about 4500 K (Saumon et al. 1994).  CIA depletes
the flux from the near-IR and pushes it into the optical.  Solar metallicity
isochrones from Dotter et al. (2008), using synthetic fluxes from PHOENIX
model atmospheres, indicate that the downturn begins at 0.6M$_{\odot}$ and 
continues at more or less constant color down below 0.1M$_{\odot}$. 

There are other prominent features in the M67 CMD. For example, 
the equal-mass binary sequence which parallels the MS in the range 
12$<$K$_S$$<$15 stands-out in the deep M67 CMD. At fainter 
magnitudes, the photometric sequence of the MS binaries blends in 
with the single star MS. In addition, the CMD shows a prominent
population of blue straggler stars and a small but significant
core-helium-burning red clump at K$_S$$\sim$8.0 and (J-K$_S)$$\sim$0.68. 

\section{Distance of M67}

The deep near-IR CMD of M67 shown in Fig. 1 allows us to derive a
new value for the distance to this important cluster. To proceed, however,
requires the adoption of a metallicity and a reddening for M67. For these,
we turn to the literature where an extensive body of work exists to
answer this question. The recent work of Taylor (2007) presents an
exhaustive analysis of published reddenings and metallicities for M67.
He finds a reddening of E(B--V) = 0.041 $\pm$ 0.004 and a metal
abundance of [Fe/H] = --0.009 $\pm$ 0.009. For the purpose of the
present paper, we will assume that the abundance of M67 is identically
the solar value. In addition, we adopt E(J--K$_S$) = 0.53 E(B--V)
from the work of Cambr\'{e}sy et al. (2002), which means that
E(J--K$_S$) = 0.022.

%
%
%

Armed with these quantities, we can proceed to fit the MS of
M67 to the field stars with {\it Hipparcos} parallaxes and 2MASS
photometry. These were assembled by Sarajedini et al. (2004) wherein
all of the relevant information is given for each star in their Table 2.
The colors of these stars must be corrected for the difference in
metallicity between M67 and that of the star. This step is accomplished
using the same technique as that described by Sarajedini et al. (2004) and
Percival et al. (2003).  To be consistent with Percival et al. (2003), we 
will use only stars with abundances in the range 
$-0.45<$[Fe/H]$< +0.35$ for the MS fitting. Furthermore, we exclude 
HIP 84164, since it is clearly an outlier in the CMDs of Percival et al. (2003). 
This leaves us with 46 field stars to be used in the MS fitting.

First, the slope of the main sequence in the absolute magnitude range
of the field stars is established using the theoretical isochrones of
Girardi et al. (2002) on the 2MASS system. 
This MS slope is then used to offset the color of each field
star along a vector with this slope to the value it would have at M$_{K_S}$ = +4.0,
which is approximately in the middle of the absolute magnitude distribution.
These transformed colors are plotted as a function of the known stellar
metallicity to produce the diagram shown in Fig. 2. A least squares fit
to these data with 2-$\sigma$ rejection yields 
(J--K$_S$) = 0.528 + 0.106[Fe/H] with a root mean square deviation of
0.022 mag in color. The slope of this relation 
($\Delta$(J--K$_S$) / $\Delta$[Fe/H])  is then applied to the colors of the
field stars in order to adjust them to an abundance of [Fe/H] = 0.0.
We note in passing that the dashed line in Fig. 2 is the result of
applying this procedure to Girardi et al. (2002) isochrones converted
to the photometric system of Bessell \& Brett (1988) suggesting
very little difference between the two systems.

We also require a fiducial sequence that best represents the location
of the M67 MS in the magnitude range of the field stars 
(i.e. 3.5 $\lea$ M$_{K_S}$ $\lea$ 4.5). For this, we began with the
proper motion selected 
deep M67 photometry shown in the upper panel of Fig. 3 and fit
a sequence of points to the MS by eye. These points were then
represented by a polynomial and offset in color by the adopted
reddening to M67 as shown by the solid line in Fig. 3. The magnitude offset 
was determined by fitting
this MS fiducial to the 46 field stars described above using the magnitude 
errors as weights. This yields a distance modulus of 
(m--M)$_{K_S}$ = 9.72 $\pm$ 0.05. The resultant fit is shown in 
the lower panel of Fig. 3. 
The absolute distance modulus is then (m--M)$_0$ = 9.70 $\pm$ 0.05.
The error on these values includes the fitting error (0.010 mag) along with 
the errors due to uncertainties in the reddening 
($\sigma$$_{E(J-K)}$ = 0.01 translates to 
$\sigma$$_{(m - M)_K}$ $\sim$ 0.035) and the metallicity value of the cluster
relative to the field stars ($\sigma$$_{[Fe/H]}$ = 0.1 dex translates 
to $\sigma$$_{(m - M)_K}$ $\sim$ 0.042). It is important to note
that the quoted uncertainty includes only these random errors; 
systematic errors such as those present in the Lutz-Kelker corrections
to the field star parallaxes are not included. In any case,
our distance for M67 is approximately in the middle of the range of 
published values as shown by Sarajedini et al. (2004). 

\section{Comparison With Theoretical Isochrones}

With values for the reddening, distance, and metallicity of M67, it is
possible to compare the theoretical isochrones to the CMD in order
to examine the robustness of the models at faint magnitude levels,
especially where the MS turns down. 
The two panels of Fig. 4 show the M67 data compared with solar
abundance isochrones from Girardi et al. (2002) and Dotter et al. (2008)
 in the 2MASS photometric system. In both cases, we have
adopted the reddening from above but adjusted the distance
modulus slightly (within the errors) to fit the main sequence
of M67 at K$_S$$\sim$14. The Girardi et al. models yield a best fit age
of 3.5 Gyr while the Dotter et al.  models yield a best fit age of 4 Gyr.

The quality of the isochrone fits is not the same between the Girardi et al.
and Dotter et al. models. In particular, the Girardi models seem to incorporate
too much core overshooting at the main sequence turnoff, thus leading to
a blueward hook that is more pronounced than the data indicates.
In addition, in the region of the lower main sequence as temperature
decreases, the Girardi models turn downward sooner than the the cluster data.
In contrast, the Dotter et al. models reproduce these features more faithfully.

The temperature scales of the two isochrones are in reasonable agreement 
near the MSTO but around 4500 K they
begin to diverge, with the Girardi models becoming considerably hotter than the
Dotter models as mass decreases along the MS (see Figure 4 of Dotter et al. 2008). 
It is this difference in the temperature scales that causes the disagreement seen in
the CMDs where the Girardi models begin to turn down at a bluer color.  The
difference itself is likely due to differences in the adopted physics, primarily the
equation of state and low-Temperature opacities (see Dotter et al. 2008, 
section 5.1 for more details).

\section{Calibration of a New Age Indicator}

Recently, Calamida et al. (2009) presented a new age indicator that makes
use of near-IR J--K colors. The diagnostic exploits the difference in color
between the MS turnoff (TO) and the point at which the MS turns down (TD)
and becomes nearly vertical [$\Delta$(J--K$_S$)$_{TD}^{TO}$]. 
As age increases, the value of
$\Delta$(J--K$_S$)$_{TD}^{TO}$ becomes smaller. 
To be useful, this age diagnostic requires calibration, which is typically
done by using theoretical isochrones. However, it is 
important to test the predictions of the isochrones in this regard by 
comparing them observational data of open clusters whose ages have been 
measured independently. In the case of the $\Delta$(J--K$_S$)$_{TD}^{TO}$
quantity, deep observations that reach the faint lower main sequence with
good precision are required. While this is a challenging prospect in
Galactic globular clusters, it is certainly a more tractable goal in
open clusters because of their relative proximity. 

We begin by exploiting the 2MASS point source catalog looking for nearby
open clusters with deep photometry. Two systems appear to have adequate
photometry for the identification of the turndown on the lower main 
sequence: NGC 2516 and M44 (NGC 2632).  To these, we add the M67
data analyzed here and deep near-IR photometry of the
old open cluster NGC 6791 (T. von Hippel et al. 2009, in preparation).
The latter were obtained with the Gemini 8-m telescope equipped with the
Near-IR Imager (NIRI) instrument. The resultant photometry has been calibrated 
to the 2MASS system using bright stars in common between the two catalogs. 

Figures 5 through 8 illustrate our technique for measuring 
$\Delta$(J--K$_S$)$_{TD}^{TO}$ in
these four open clusters. First, we isolate the MS stars that likely belong to the
cluster so as to minimize contamination from field stars. This is done
by eye with some knowledge of where we expect the main sequence
of the cluster to be located. The likely MS stars are shown as the larger
points in the lower panels of Figs. 5 through 8. The upper panels of these 
figures display the color histograms of the likely cluster members in each case.
The bluest peak in this distribution is the MS turnoff [(J--K$_S$)$_{TO}$]
while the reddest one represents the MS turndown [(J-K$_S$)$_{TD}$]. 
Gaussian fits to these peaks then yield the
adopted values for these quantities, which are given in Table 1. Also
listed there are the cluster metallicities and ages from their respective 
sources. We adopt an uncertainty of 0.1 dex in the
logarithm of the ages so that the error in the age of NGC 2516 is
$\sim$40 Myr while the error in the age of NGC 6791 is $\sim$1 Gyr.

It is clear from Fig. 5 that determining the MS turnoff color
of NGC 2516 is the most problematic because of its sparseness.
In this case, we limit ourselves to stars with K$_S$$<$9 and derive
the histogram shown
by the thick dashed line, which is referred to the right-hand ordinate
in the upper panel of Fig. 5. This then facilitates the Gaussian fit to
determine the TO color for NGC 2516.

Figure 9 shows the variation of $\Delta$(J--K$_S$)$_{TD}^{TO}$ with 
age for the four clusters
in our sample. As expected, the quantity $\Delta$(J--K$_S$)$_{TD}^{TO}$ 
decreases as
a cluster gets older. A weighted least squares fit to these data (dashed line
in the upper panel of Fig. 9) yields
$\Delta$(J--K$_S$)$_{TD}^{TO}$ = (3.017 $\pm$ 0.347) -- (0.259 $\pm$ 0.037)*Log Age (yrs). Thus to achieve a precision of better than 0.1 in the logarithm 
of the age ($\sim$20\% error), the quantity $\Delta$(J--K$_S$)$_{TD}^{TO}$  
would need to be measured with an error of less than 0.026 mag.

The lines in the lower panel of Fig. 9 represent the predicted 
behavior based on the Dotter et al. (2008) and Girardi et al. (2002)
isochrones for the indicated 
metallicities. It is clear that, given the uncertainties in the observational 
quantities, the model predictions are consistent with the data.

While Fig. 9 indicates consistency between model predictions of the relationship
between age and $\Delta$(J--K$_S$), it is important to consider the 
limitations and uncertainties in these predictions.  The
sensitivity of $\Delta$(J--K$_S$) decreases with increasing age.
As such, the method is likely to be of more use for measuring the ages of young
to intermediate age open clusters.  Since these systems are also the ones most
likely to have sufficiently deep CMDs, this issue is not a major concern.
The $\Delta$(J--K$_S$) method may prove useful for extracting ages
of young open clusters in highly obscured fields where standard isochrone fitting
to the CMD is complicated by significant field star contamination.

Dotter et al. (2008) discuss some of the major differences that can be 
found among different sets of isochrones for ages and metallicities 
appropriate for open clusters considered in this paper (see their 
section 5.1 and Figures 2-5). Particularly important 
to this discussion are the equation of state--especially for the lower MS and 
the location of the MS turndown, the treatment of convective core 
overshoot--for the location of the MSTO at a given age, and the adopted 
color-effective temperature transformations--for the 
morphology of the isochrones in the J--K$_S$ CMD. Other factors, such 
as the adopted solar metallicity and $\Delta$Y/$\Delta$Z, are relevant 
as well.  Finally, we argue that given its insensitivity to age, the location 
of the MS turndown in the J--K$_S$ CMD
provides an excellent calibration for theoretical models: both the effective 
temperature scale of the stellar evolution models and the color-effective 
temperature transformations.

\section{Summary and Conclusions}

We present an analysis of the 2MASS calibration 
photometry of the old open cluster M67 (NGC 2682). While the 2MASS
PSC has seen broad appeal and utilization, the calibration data has not
been widely analyzed. Yet, the M67 CMD resulting from these data
is more than 3 magnitudes deeper than the one from the PSC extending
to more than 7 magnitudes below the main sequence turnoff of M67.
From this diagram, we draw the following conclusions:

\noindent 1) Literature value suggest a mean reddening of
E(B--V) = 0.041 $\pm$ 0.004 and a metal abundance of 
[Fe/H] = --0.009 $\pm$ 0.009. Using these quantities along with
2MASS photometry and {\it Hipparcos} astrometry for 46 field stars
of known metallicity, we perform main sequence fitting on M67 to 
determine distance moduli of
(m--M)$_{K_S}$ = 9.72 $\pm$ 0.05 and  
(m--M)$_0$ = 9.70 $\pm$ 0.05, which are consistent with published
values. Note that the quoted distance uncertainty does not include 
contributions due to systematic errors such as those associated with
the Lutz-Kelker corrections.

\noindent 2) We compare the theoretical isochrones of Girardi et al. (2002)
and Dotter et al. (2008) to the 2MASS CMD of M67. Generally speaking, both sets
of models reproduce the morphology of the CMD within two magnitudes
of the MSTO and brighter. However, the Dotter et al. models provide a more faithful 
representation of the of lower MS, including the color of the MS turndown.
These comparisons suggest an age of 3.5 (Girardi) or 4.0 (Dotter) Gyr for M67. 

\noindent 3) We combine the 2MASS calibration data for M67, PSC
photometry for NGC 2516 and M44, and our own data for the 
old metal-rich open cluster NGC 6791 in order to calibrate the
newly devised age diagnostic of Calamida et al. (2009). The
diagnostic  - the difference in
(J--K$_S$) color between the main sequence turnoff (TO) and the point on the 
lower main sequence where it turns down (TD) and becomes nearly vertical
[$\Delta$(J--K$_S$)$_{TD}^{TO}$] - becomes smaller for older clusters.
We find 
$\Delta$(J--K$_S$)$_{TD}^{TO}$ = (3.015 $\pm$ 0.347) -- (0.259 $\pm$ 0.037)*Log Age (yrs) with a very small dispersion in the fitted points.

\acknowledgements

We thank Mike Skrutskie for bringing the deep calibration dataset of M67
to our attention.
We are grateful to Ted von Hippel for helpful comments on this manuscript. 
We also thank an anonymous referee for helpful suggestions that greatly
improved the presentation of this paper.
A. S. was supported by grant AST - 0606703 from the National
Science Foundation. A. K. was partially supported by a
Research Experiences for Undergraduates supplemental grant
from the National Science Foundation. A. D. acknowledges support
from the Canadian Institute for Theoretical Astrophysics and the 
Natural Sciences and Engineering Research Council of Canada.


\begin{deluxetable}{lcccccc}
\tablecaption{Open Clusters}
\tabletypesize{\scriptsize}
\tablewidth{0pt}
\tablehead{
\colhead{Name}
  &\colhead{[Fe/H]}
  &\colhead{Log Age (yrs)}
  &\colhead{Age Reference}
  &\colhead{$(J-K_S)_{TO}$}
  &\colhead{$(J-K_S)_{TD}$}
  &\colhead{$\Delta$(J--K$_S$)$_{TD}^{TO}$}
}
\startdata
%
NGC 2516       & $+0.06 \pm 0.03^a$  & 8.15 & Terndrup et al. (2002)& --0.002 $\pm$ 0.037 & 0.900 $\pm$ 0.061 & 0.902 $\pm$ 0.071 \\
NGC 2632 (M44) &$+0.057\pm0.022^b$&  8.78 & Claver et al. (2001)  & ~~0.114 $\pm$ 0.041 & 0.866 $\pm$ 0.027 & 0.752 $\pm$ 0.049 \\
NGC 2682 (M67) &$-0.009\pm0.009^c$& 9.60 & Sarajedini (1999)     & ~~0.328 $\pm$ 0.028 & 0.856 $\pm$ 0.020 & 0.528 $\pm$ 0.034 \\
NGC 6791       &$+0.39\pm0.05^d$  & 9.98 & Sarajedini (1999)     & ~~0.451 $\pm$ 0.023 & 0.888 $\pm$ 0.029 & 0.437 $\pm$ 0.037 \\
\enddata
\tablerefs{$a$--Twarog et al. (1997); $b$--Taylor (2008); $c$--Taylor (2007); $d$--Carraro et al. (2006)}
\end{deluxetable}



\clearpage
\begin{figure}
\epsscale{0.95}
\plotone{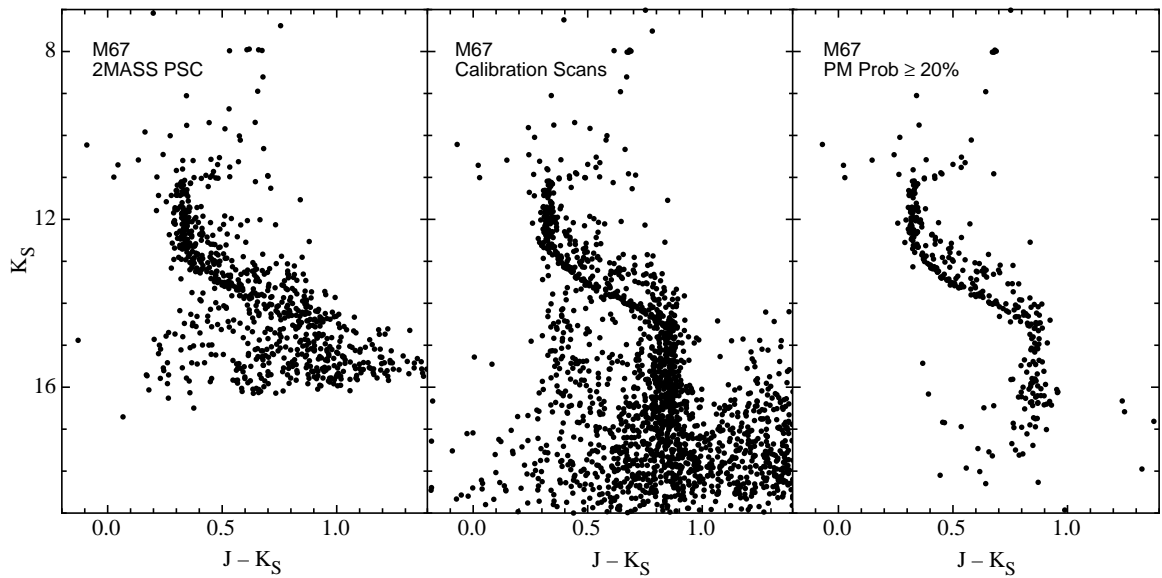}
\caption{The color-magnitude diagram of M67 from the 2MASS point source
catalog (left panel) and the Combined Calibration database (middle panel). The
right panel also shows the Combined Calibration data but including only stars 
with a $\geq$20\% probability of membership based on the proper motions of 
Yadav et al. (2008).}
\end{figure}

\clearpage
\begin{figure}
\epsscale{0.95}
\plotone{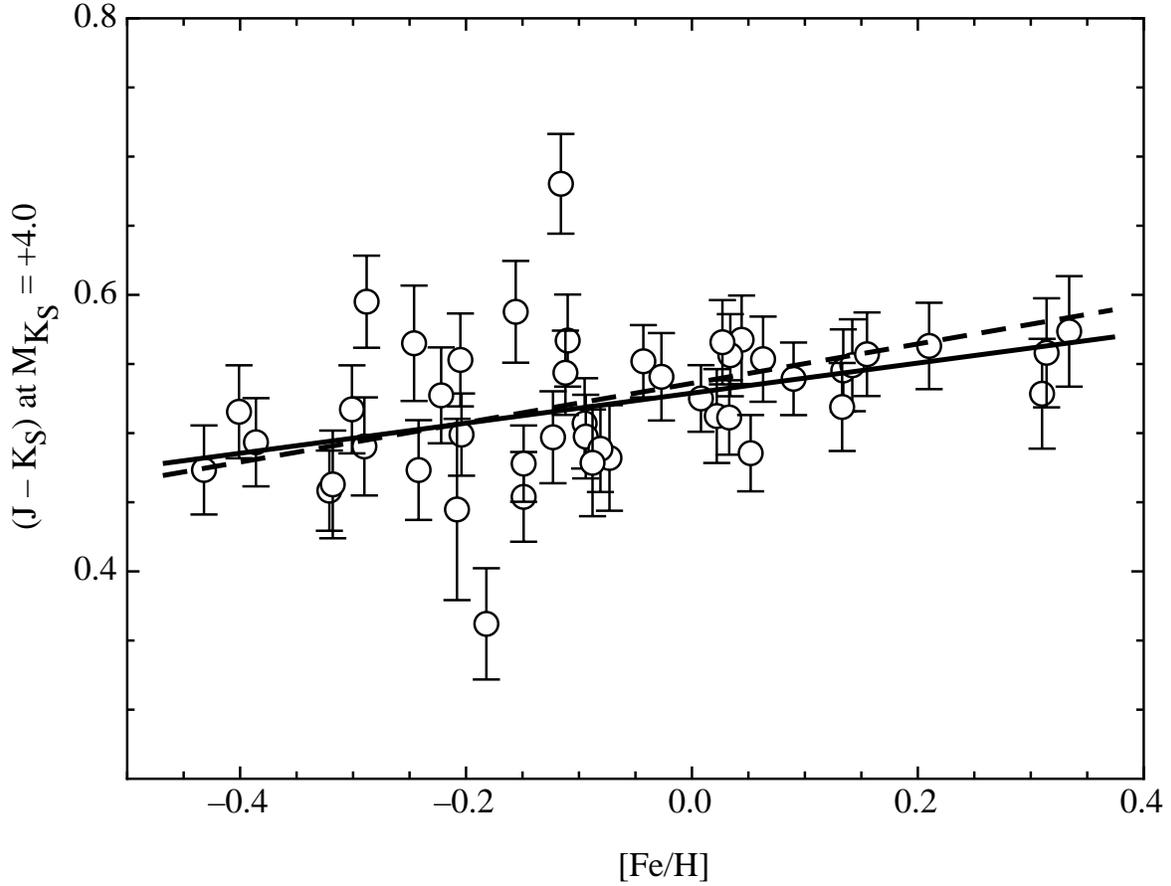}
\caption{The color each field star would have at M$_{K_S}$ = +4.0 as a function
of the metal abundance for 46 stars with --0.45$<$[Fe/H]$<$+0.35 from the
work of Sarajedini et al. (2004). The solid line is the least squares fit with
2-$\sigma$ rejection applied. The slope of the fitted line is 
$\Delta(J-K)$ / $\Delta[Fe/H]$ = 0.106. For comparison, the dashed line
represents the result of performing this experiment after transforming
the 2MASS photometry to the system of Bessel \& Brett (1988) suggesting
very little difference between the two systems.}
\end{figure}

\clearpage
\begin{figure}
\epsscale{0.7}
\plotone{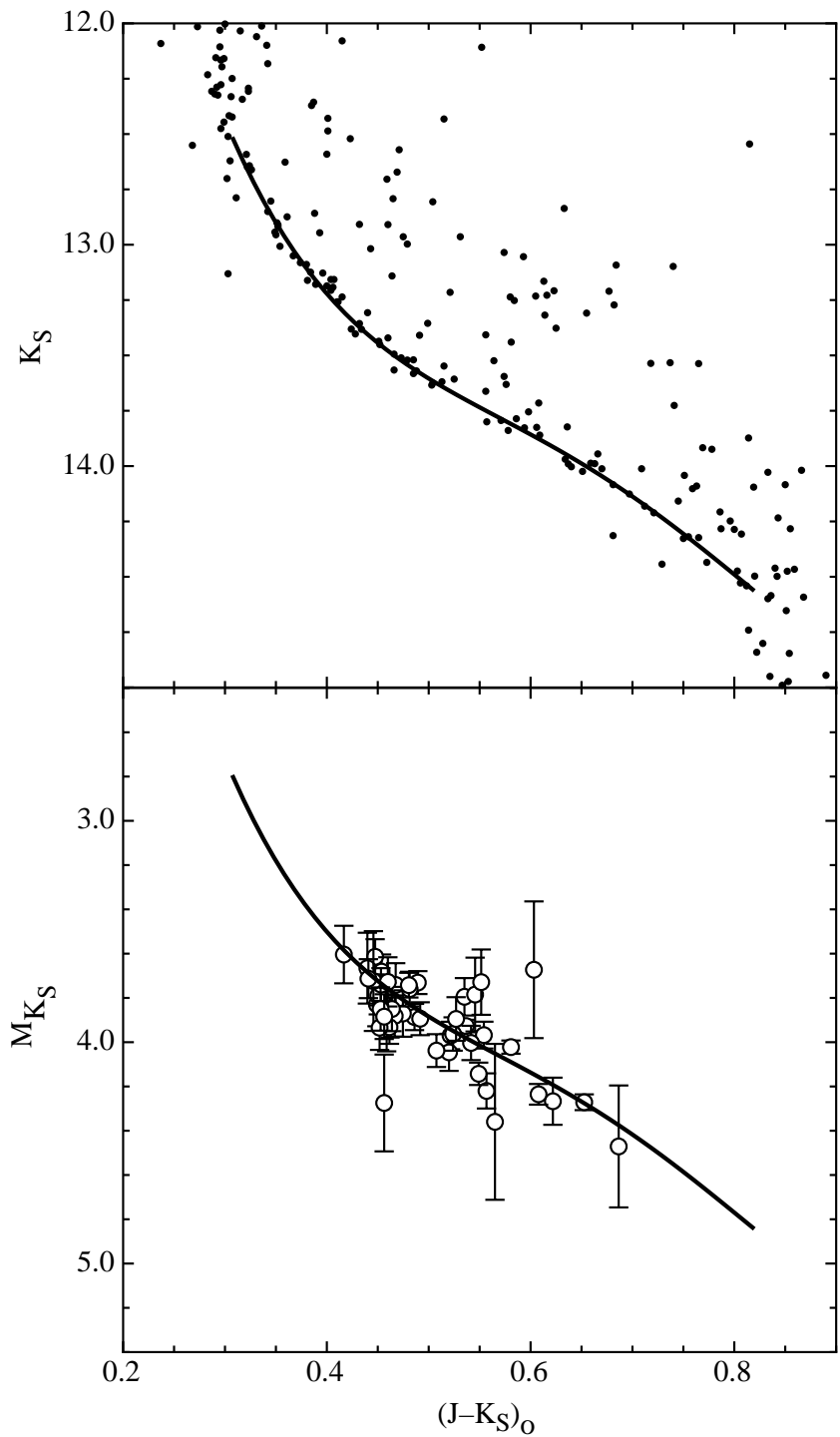}
\caption{The upper panel shows the proper motion selected
deep M67 2MASS photometry (filled
circles) corrected for a reddening of E(B-V) = 0.041. The solid line is the adopted 
fiducial for the main sequence. 
The open circles in the lower panel are the field stars with {\it Hipparcos}
parallaxes and 2MASS photometry corrected for distance. The solid line
is the M67 fiducial from the upper panel fitted to the field stars resulting in
an inferred distance of (m--M)$_{K_S}$ = 9.72. }
\end{figure}

\clearpage
\begin{figure}
\epsscale{0.95}
\plotone{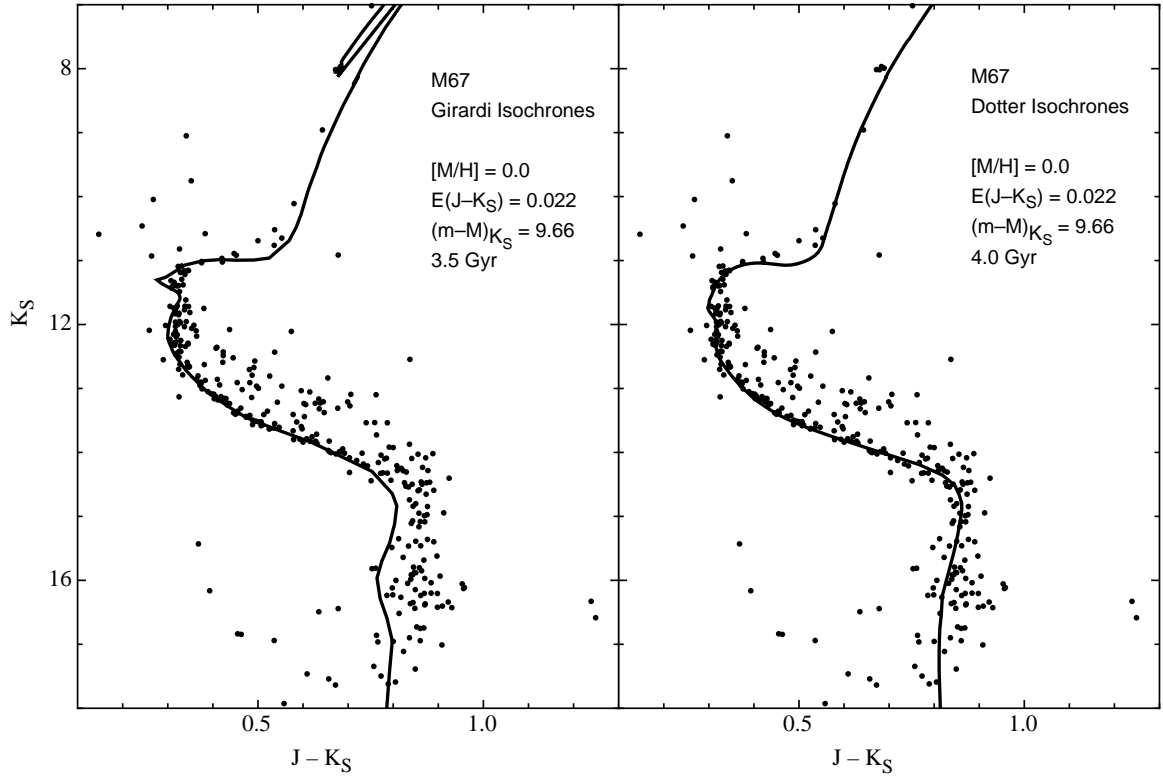}
\caption{The color-magnitude diagram of M67 from the proper motion
selected 2MASS Combined 
Calibration database compared with theoretical models for given values of 
reddening, distance, and metallicity. The left panel shows the 3.5 Gyr
isochrone from Girardi et al. (2002). The right panel displays the
4.0 Gyr track from Dotter et al. (2008).}
\end{figure}

\clearpage
\begin{figure}
\epsscale{0.55}
\plotone{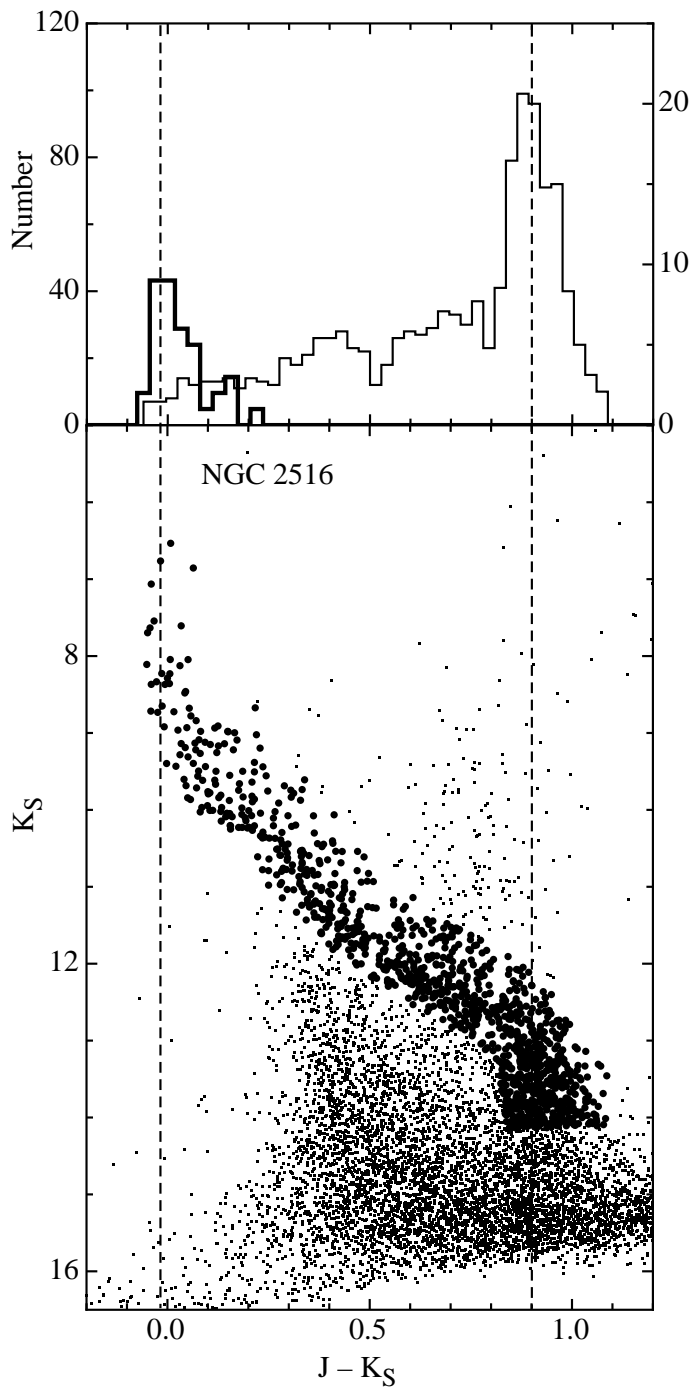}
\caption{The lower panel shows the color-magnitude diagram of NGC 2516 from 
the 2MASS Point Source Catalog. The larger points denote
the stars used to construct the color function shown in the upper panel. Because
of the sparseness of the main sequence turnoff region, we further
limit the color function in this region to stars with K$_S$$<$9. Note
that this latter histogram is referred to the right hand ordinate axis. The
dashed lines represent the locations of the main sequence turnoff color and 
the turndown color (see text).}
\end{figure}

\clearpage
\begin{figure}
\epsscale{0.55}
\plotone{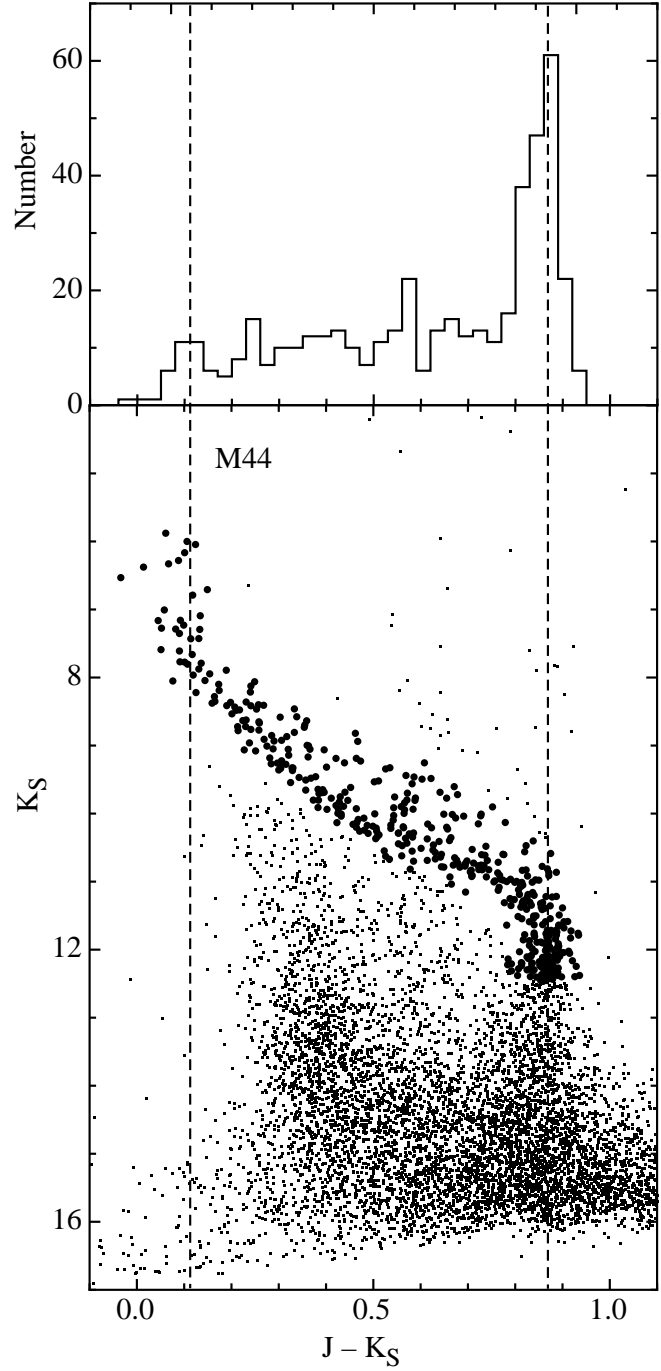}
\caption{Same as Fig. 5 except that the plotted data are the M44 photometry
from the 2MASS Point Source Catalog.}
\end{figure}

\clearpage
\begin{figure}
\epsscale{0.55}
\plotone{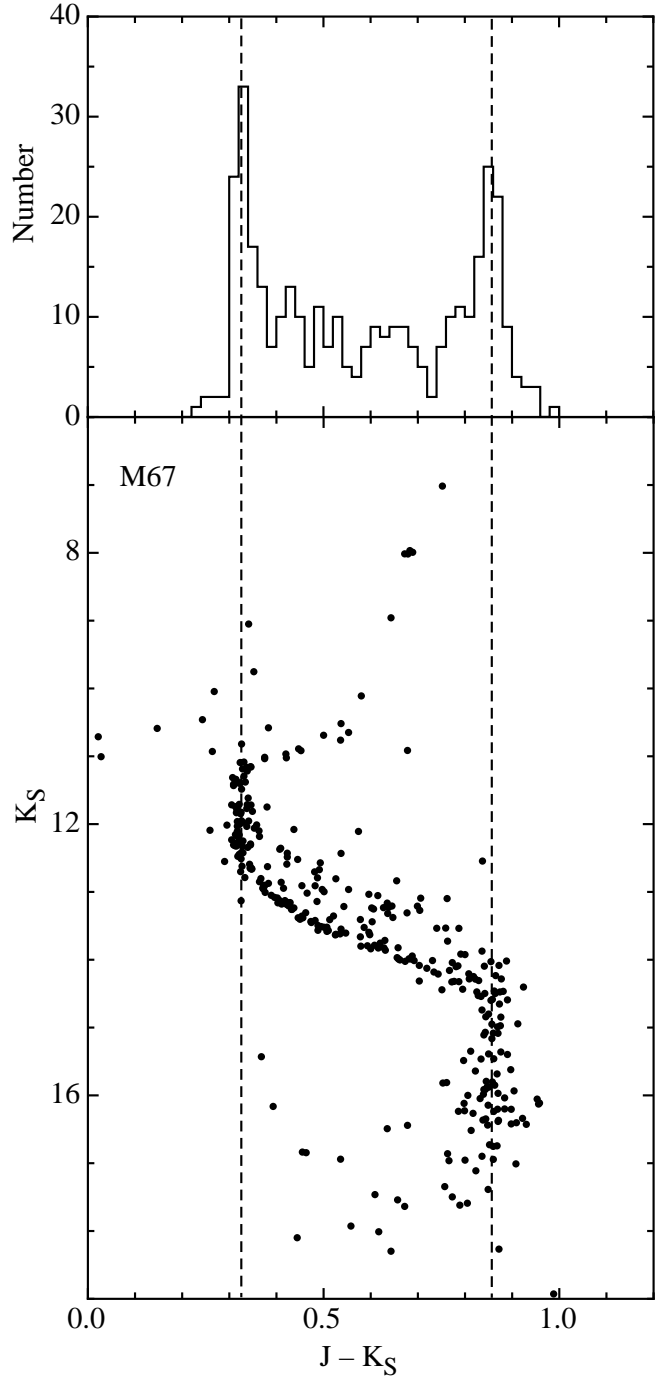}
\caption{Same as Fig. 5 except that the plotted data are the M67 photometry
from the proper motion selected 2MASS Combined Calibration database.}
\end{figure}

\clearpage
\begin{figure}
\epsscale{0.55}
\plotone{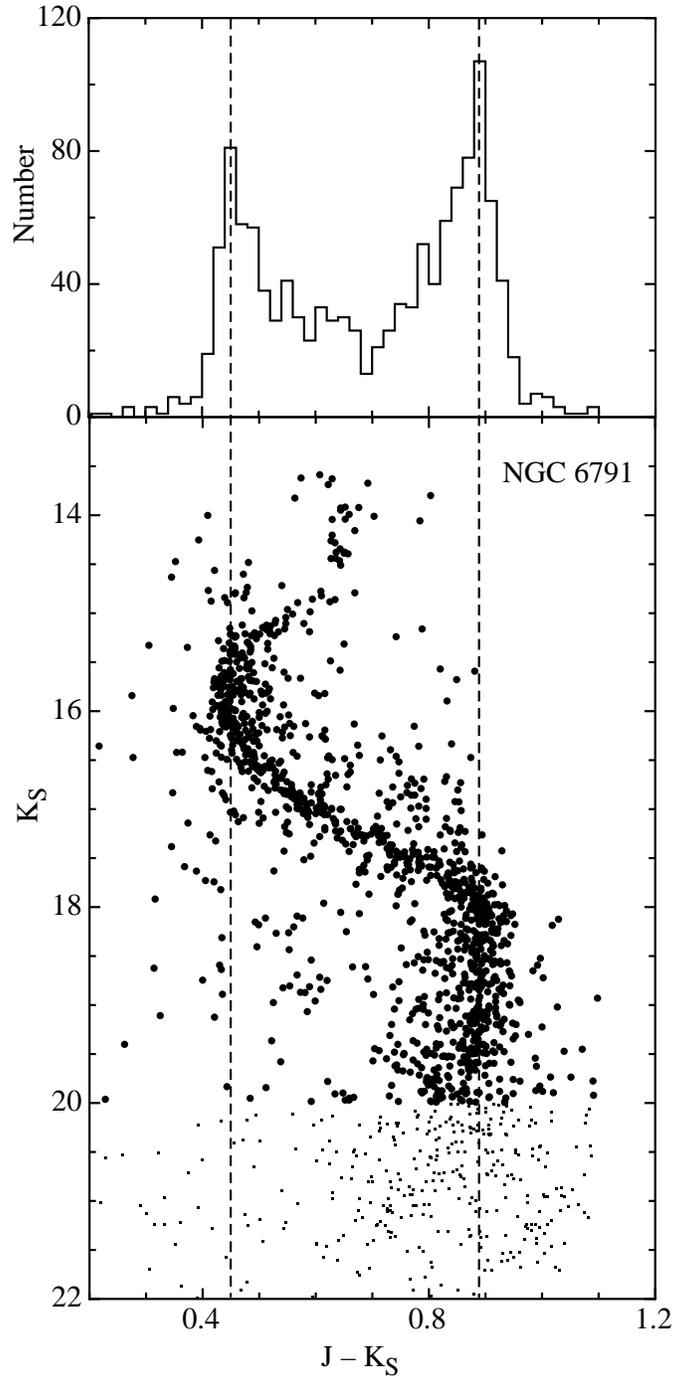}
\caption{Same as Fig. 5 except that the plotted are the
NGC 6791 photometry from von Hippel et al. (2009, in preparation).}
\end{figure}

\clearpage
\begin{figure}
\epsscale{0.7}
\plotone{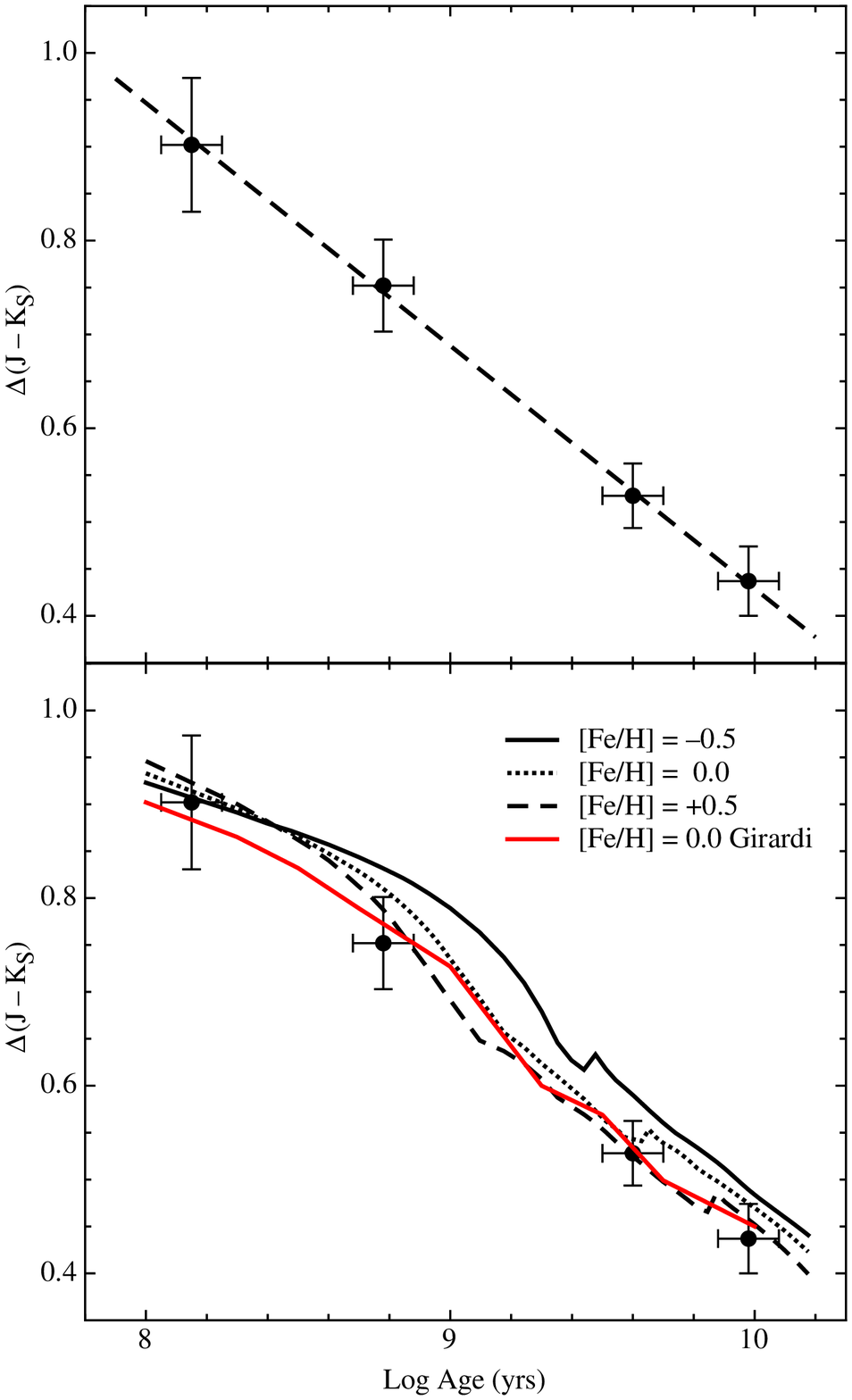}
\caption{The upper panel shows the variation of 
$\Delta$(J--K$_S$)$_{TD}^{TO}$  with the 
logarithm of the age in years for the clusters in Table 1. The dashed line
is the weighted least squares fit to the data. The lower panel displays
the same data points but this time compared with the theoretical models
of Dotter et al. (2008, black lines) and Girardi et al. (2002, red line).}
\end{figure}


\begin{thebibliography}{}
\bibitem[]{} An, D., Terndrup, D. M., Pinsonneault, M. H. 2007, \apj,
671, 1640
\bibitem[]{} Bessell, M. S. \& Brett, J. M. 1988, \pasp, 100, 1134
\bibitem[]{} Calamida, A., et al. 2009, in The Ages of Stars, in press
\bibitem[]{} Cambr\'{e}sy, L., Beichman, C. A., Jarrett, T. H., \&
Cutri, R. M. 2007, AJ, 123, 2559
\bibitem[]{} Claver, C. F., Liebert, J., Bergeron, P., \& Koester, D.
2001, \apj, 563, 987
\bibitem[]{} Dotter, A., Chaboyer, B. Jevremovi\'c, D., Kostov, V., Baron, E., \& Ferguson, J. W. 2008, \apjs, 178, 89
\bibitem[]{} Girardi, L., Bertelli, G., Bressan, A., Chiosi, C., 
Groenewegen, M. A. T., Marigo, P., Salasnich, B., \& Weiss, A.,
2002, \aap, 391, 195
\bibitem[]{} Grocholski, A. \& Sarajedini, A. 2002, \aj, 123, 1603
\bibitem[]{} Nikolaev, S., Weinberg, M. D., Skrutskie, M. F., Cutri, R. M.,
Wheelock, S. L., Gizis, J. E., \& Howard, E. M. 2000, 120, 3340
\bibitem[]{} Percival, S. M., Salaris, M., \& Kilkenny, D. 2003, \aap, 400, 541 
\bibitem[]{} Sarajedini, A. 1999, \aj, 118, 2321
\bibitem[]{} Sarajedini, A., Brandt, K., Grocholski, A. J., \& Tiede, G. P.
2004, \aj, 127, 991
\bibitem[]{} Saumon, D. 1994, in ``The equation of state in astrophysics,"
Proceedings of IAU Colloquium No. 147 (Cambridge: Cambridge University 
Press), edited by G. Chabrier and E. Schatzman, Evry, p.306
\bibitem[]{} Skrutskie, M. et al. 2006, \aj, 131, 1163
\bibitem[]{} Taylor, B. J. 2007, \aj, 133, 370
\bibitem[]{} Terndrup, D. M., Pinsonneault, M., Jeffries, R. D., Ford, A., Stauffer, J. R., \& Sills, A. 2002, \apj, 576, 950
\bibitem[]{} Yadav, R. K. S., Bedin, L. R., Piotto, G., Anderson, J., Cassisi, S.,
Villanova, S., Platais, I., Pasquini, L., Momany, Y., \& Sagar, R. 2008, \aap, 484, 609
\end{thebibliography}
\end{document}